\documentclass[epj]{webofc}
\usepackage[varg]{txfonts}   
\woctitle{MESON2014 - the 13$^\textrm{th}$ International Workshop on Meson Production, Properties and Interaction}
\begin{document}
\selectlanguage{english}
\begin{flushright}
{\sf  MITP/14-014 \\
  } 
\end{flushright}
\vspace{-0.5cm}
\title{$\eta$ and $\eta'$ transition form factors from Pad\'e approximants}

\author{Pablo Sanchez-Puertas\inst{1}\fnsep\thanks{\email{sanchezp@kph.uni-mainz.de}. Supported by the Deutsche Forschungsgemeinschaft DFG through the Collaborative Research Center ``The Low-Energy Frontier of the Standard Model" (SFB 1044) and by the PRISMA Cluster of Excellence.} \and
        Pere Masjuan\inst{1} 
}

\institute{PRISMA Cluster of Excellence, Institut f\"ur Kernphysik, \\ Johannes Gutenberg-Universit\"at, Mainz D-55099, Germany 
          }

\abstract{%
  We employ a systematic and model-independent method to extract, from space- and time-like data, the $\eta$ and $\eta'$ transition form factors (TFFs) obtaining the most precise determination 
  for their low-energy parameters and discuss the $\Gamma_{\eta\rightarrow\gamma\gamma}$ impact on them. 
  Using TFF data alone, we also extract the $\eta-\eta'$ mixing parameters, which are compatible to those obtained from more sophisticated and input-demanding procedures.
}
\maketitle
\section{Introduction}
\label{intro}

The hadronic structure of neutral pseudoscalar mesons may be probed via the two-photon mechanism. The most general matrix element for such process is given by
\begin{equation}
{\mathcal{M}}_{P\gamma^*\gamma^*} = ie^2\varepsilon^{\mu\nu\rho\sigma}q_{1,\mu}\epsilon_{1,\nu}q_{2,\rho}\epsilon_{2,\sigma}F_{P\gamma^*\gamma^*}(q_1^2,q_2^2),
\end{equation}
where $q_i(\epsilon_i)$ stands for the i-\textit{th} photon momentum (polarization) and $F_{P\gamma^*\gamma^*}(q_1^2,q_2^2)$ is the pseudoscalar transition form factor (TFF) encoding all the strong-interaction 
effects. Of particular interest is the single virtual TFF $ F_{P\gamma^*\gamma}(Q^2) \equiv F_{P\gamma^*\gamma^*}(-q^2,0)$ for which many measurements are available. At low energies, the TFF can be expressed 
in terms of its low-energy parameters (LEPs) $b_P,c_P,d_P,...$
\begin{equation}
F_{P\gamma^*\gamma}(Q^2) = F_{P\gamma\gamma}(0)\left( 1 - b_P\left(\frac{Q^2}{m_P^2}\right) + c_P\left(\frac{Q^4}{m_P^2}\right) - d_P\left(\frac{Q^6}{m_P^2}\right) + ...\right).
\label{eq-5}
\end{equation} 
However, due to the non-perturbative behavior of QCD at low energies, neither the TFF, nor its LEPs, can be 
calculated from first principles. Only its low- and high-energy limits are known from the axial anomaly~\cite{Adler:1969gk} and perturbative QCD~\cite{Lepage:1980fj}, respectively. Remarkably, 
both limits depend on the same parameters. In this work~\cite{Escribano:2013kba}, we focus on the $\eta$ and $\eta'$ TFFs. Using the flavor basis to describe the $\eta-\eta'$ mixing, these limits 
read~\cite{Feldmann:1998yc}
\begin{align}
F_{\eta\gamma\gamma}(0)=\left((\hat c_q/F_q)\cos\phi-(\hat c_s/F_s)\sin\phi\right)/4\pi^2\ , \label{eq-1}\\
F_{\eta'\gamma\gamma}(0)=\left((\hat c_q/F_q)\sin\phi+(\hat c_s/F_s)\cos\phi\right)/4\pi^2\ , \label{eq-2}\\
\lim_{Q^2\to\infty }Q^2F_{\eta\gamma^*\gamma}(Q^2) =2(\hat c_q F_q\cos\phi-\hat c_s F_s\sin\phi)\ , \label{eq-3}\\
\lim_{Q^2\to\infty }Q^2F_{\eta^\prime\gamma^*\gamma}(Q^2)=2(\hat c_q F_q\sin\phi+\hat c_s F_s\cos\phi) \ , \label{eq-4}
\end{align}
where $\hat c_q (\hat c_s)= 5/3 (\sqrt{2}/3)$, $F_{q(s)}$ are decay constants, and $\phi$ is the mixing angle~\cite{Feldmann:1998yc,Feldmann:1998vh,Escribano:2005qq,Escribano:2013kba}. As an attempt to 
achieve a unified description for the whole energy regime, vector meson dominance (VMD) models, which find inspiration in the 
large-$N_c$ limit of QCD, have been extensively 
used. However,  these models contain potential systematic errors coming from simplifying assumptions and large-$N_c$ corrections, which should not be ignored when calculating 
precision observables such as the hadronic light by light contribution to the $(g-2)_{\mu}$. 
This uncertainty may be observed when comparing the different determinations from space-like (SL) and time-like (TL) data for $b_{\eta}$. 
The result, which is obtained after a fit to data using the most simple VMD parametrization $F_{\eta\gamma^*\gamma}(Q^2) = F_{\eta\gamma^*}(0)/(1+Q^2/\Lambda^2) $, is significantly different when 
using SL or TL data. Such result may be taken as the crudest one in a systematic expansion in terms of Pad\'e approximants (PA) as suggested in Ref.~\cite{Masjuan:2008fv}. Only 
when taking into account the systematic error from this expansion, the different determinations agree. In this work~\cite{Escribano:2013kba}, we extend the PA description for 
the $\pi^0$-TFF in Ref.~\cite{Masjuan:2012wy} to the $\eta$ and $\eta'$ cases.

\section{Method and results for the TFF}
\label{sec-1}

Pad\'e approximants $P^N_M(x)$ are rational functions of two polynomials $P^N_M(x)=R_N(x)/Q_M(x)$ of degree $N$ and $M$ respectively, which coefficients are related to the original 
function $f(x)$ to be approximated through the condition $f(x)-P^N_M(x) = \mathcal{O}(x^{N+M+1})$~\cite{Baker}.
They are known to converge for meromorphic and Stieltjes functions~\cite{Baker}, which has proven useful in QCD. For applications, see~\cite{Masjuan:2009wy} and references therein.\\

In our case of study~\cite{Escribano:2013kba}, the $Q^2$-dependence for $F_{P\gamma^*\gamma}(Q^2)$ as well as its analytic structure is unknown, and therefore, convergence theorems cannot be applied. 
Instead, we check the excellent performance of PA for different well-motivated physical models. 
Furthermore, since the LEPs are unknown, we extract them from a fitting procedure to the published $Q^2F_{\eta(\eta')\gamma^*\gamma}(Q^2)$ data using 
sequences of PAs. Particularly, we use the $P^N_1$ and $P^N_{N}$ sequences, from which, the VMD $(N=1)$ parametrization, is the crudest approximation. 
Having a finite amount of data, these sequences must be truncated at some finite $N$. The systematic error this implies for the LEPs determination is estimated from the models. 
Averaging over the different sequences and including this last error allows for a model-indepedent determination of the LEPs, which may be used later to systematically reconstruct the TFF 
through the use of PAs.  
\\

\begin{figure}[ht]
\centering
\includegraphics[width=4.5cm,clip]{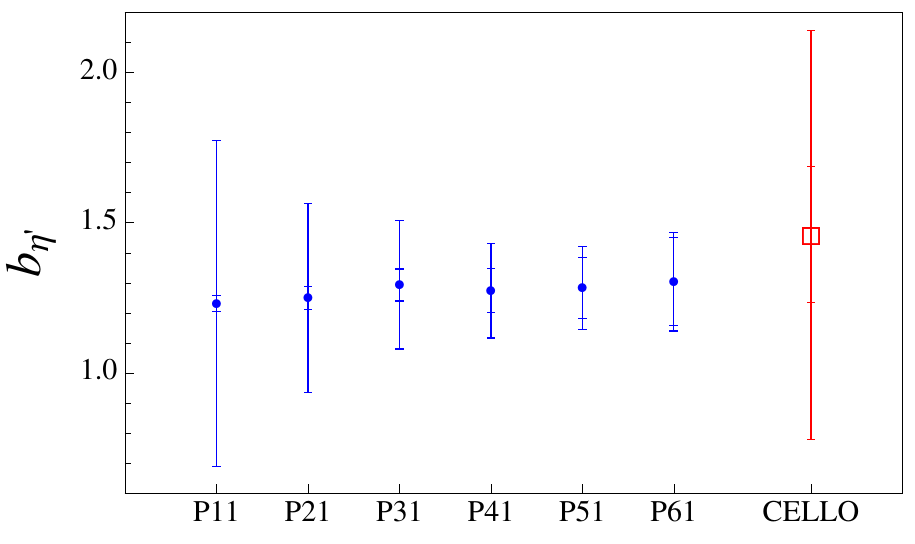}
\includegraphics[width=4.5cm,clip]{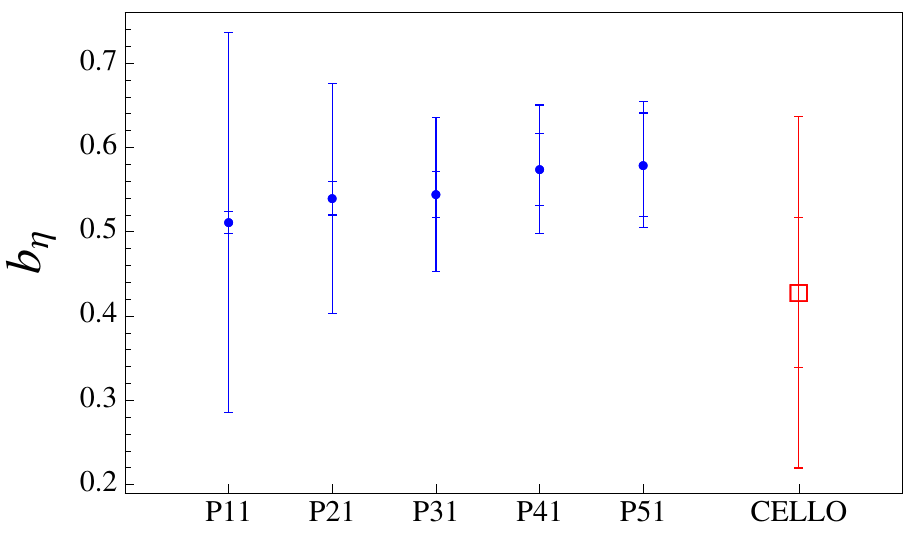}
\includegraphics[width=4.5cm,clip]{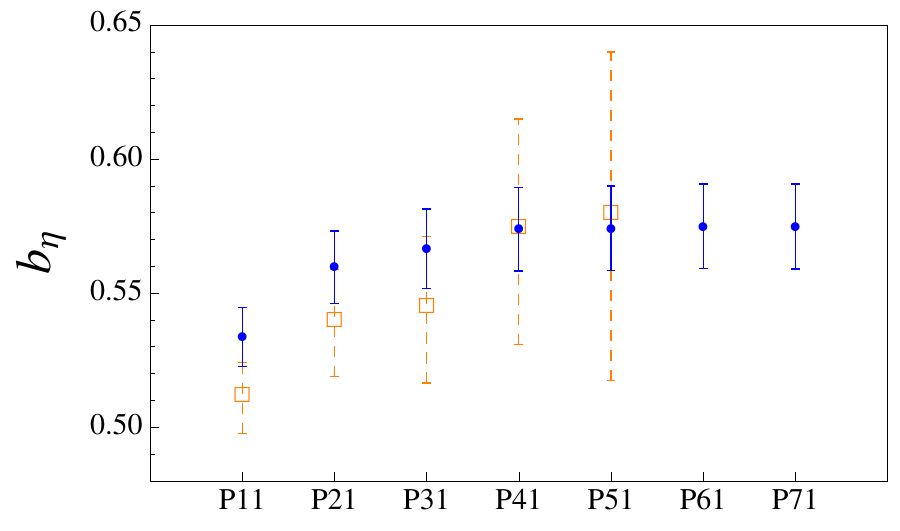}
\caption{Slope $(b_P)$ determination from the $P^N_1$ sequence. The left and central panels show the results for $\eta'$ and $\eta$ from SL data with stat. (inner band) and combined stat. and syst. (outer band) errors. The right panel shows the $\eta$ SL+TL result (solid line) together with the old SL result (dashed line). Only stat. errors are shown.}
\label{fig-1}       
\end{figure}

Our results for the LEPs from a fit to the available SL data are shown in Tab.~\ref{tab-1}, while the slope $(b_P)$ convergence pattern is illustrated in Fig.~\ref{fig-1}. Due to the systematic error, only the first 
two LEPs determination are meaningful. Our final result for $b_{\eta'}$ is the most precise to date and that of $b_{\eta}$ is comparable to the most precise experimental extraction obtained  by A2 Coll. 
$b_{\eta}=0.585(54)$~\cite{Aguar-Bartolome:2013vpw} based on low-energy TL data. A detailed comparison to different results may be found in Ref.~\cite{Escribano:2013kba}.\\
  \begin{table}
  \centering
    \caption{LEPs extraction from a fit to SL data using $P^N_1$ and $P^N_{N}$ sequences.}
    \label{tab-1}     
    \begin{tabular}{lllllllll}
      \hline
      \multicolumn{5}{c}{$\eta$ TFF} & \multicolumn{4}{c}{$\eta'$ TFF} \\ 
              & $N$  & $b_{\eta}$  & $c_{\eta}$ & $\chi^2/$dof & $N$ & $b_{\eta'}$ & $c_{\eta'}$ & $\chi^2/$dof  \\\hline
      $P^N_1$ &  $5$ & $0.58(6)$   & $0.34(8)$  & $0.80$       & $6$ & $1.30(15)$  &  $1.72(47)$ & $0.70$ \\
      $P^N_N$ &  $2$ & $0.66(10)$  & $0.47(15)$ & $0.77$       & $1$ & $1.23(3)$   &  $1.52(7)$  & $0.67$\\
       Final  &      &  $0.60(6)$  & $0.37(10)$ &              &     & $1.30(15)$  &  $1.72(47)$ & \\\hline
    \end{tabular}
  \end{table}
%
%

In Ref.~\cite{Escribano:2013kba}, we suggested that our results may be used to predict the low-$q^2$ TL region, accessible through the $P\rightarrow\gamma^*\gamma\rightarrow\overline{\ell}\ell$ Dalitz decay. 
Such measurement was recently performed by the A2 Coll. for $P=\eta$~\cite{Aguar-Bartolome:2013vpw}, and found an excellent agreement with our prediction. Their results
encouraged us to include the TL data in our fitting procedure~\cite{new}. Our new determination for the LEPs is shown in Tab.~\ref{tab-2}. The advantages of including 
the TL data are clear: we reduce significantly the systematic errors by going to higher PAs, being able to obtain up to the third derivative $d_{\eta}$, and we improve both, on convergence 
(reducing the systematic error), and statistical errors as shown in Fig.~\ref{fig-1}, right panel.
\begin{table}[h]
\centering
\caption{New determination of the $\eta$ LEPs from a fit to SL+TL data~\cite{new}.}
\label{tab-2}     
\begin{tabular}{llllll}
\hline
                & $N$ & $b_{\eta}$  &  $c_{\eta}$ & $d_{\eta}$ & $\chi^2/$dof \\\hline
$P^N_1$ &  $7$ & $0.575(16)$   &  $0.338(22)$  & $0.198(21)$  & $0.7$ \\
$P^N_N$ &  $2$ & $0.576(15)$  &  $0.340(20)$  & $0.201(19)$  & $0.7$\\
    Final    &         &  $0.576(11)$  &  $0.339(15)$  & $0.200(14)$   & \\\hline
\end{tabular}
\end{table}

Our results in Tab.~\ref{tab-2} are, by far, the most precise to date. Particularly, we believe that the precision achieved for $b_{\eta}$ will be hard to improve even if new data becomes available. Nevertheless, it 
must be stressed that the values obtained mildly depend on $F_{\eta\gamma\gamma}(0,0)$. For instance, if we would have used the value implied by the Primakoff  
$\Gamma_{\eta\rightarrow\gamma\gamma}$ decay width omitted in the PDG average~\cite{Agashe:2014kda}, we would find $b_{\eta}=0.57(6)$ and $b_{\eta}=0.570(13)$ for the SL an SL+TL extractions 
respectively, which is relevant at the obtained precision. Therefore, clarifying this experimental situation would be important.\\

%


\section{$\eta-\eta'$ mixing parameters}

\begin{figure}[ht]
\centering
\includegraphics[width=\linewidth]{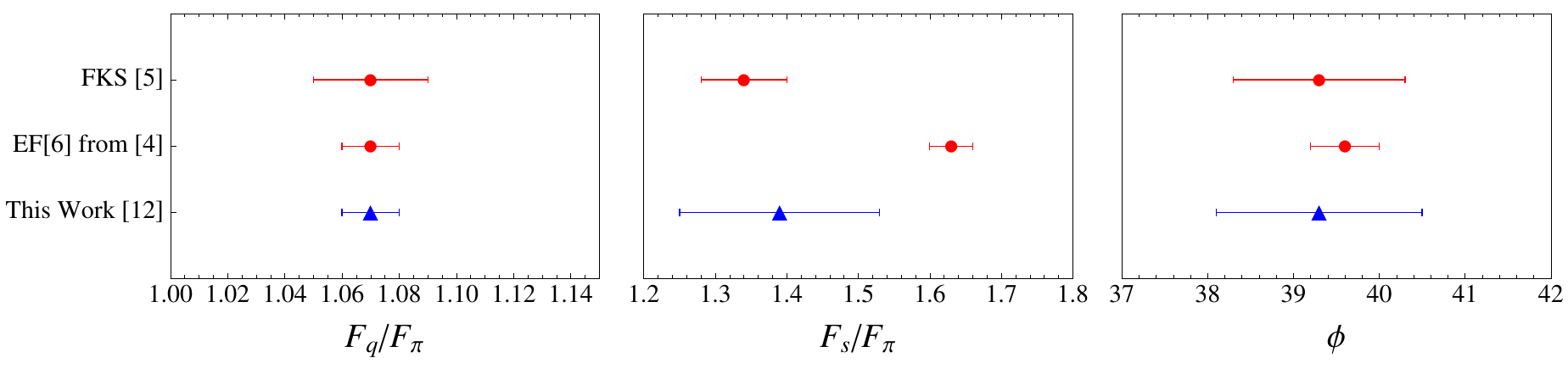}
\caption{Our mixing parameters determination (triangles) compared to other phenomenological results (circles).}
\label{fig-4}       
\end{figure}

Using the $P^N_{N}(Q^2)$ sequence results, which has the correct asymptotic behavior implemented, we can extract the TFF asymptotic value~\eqref{eq-3}. With this information, as well as 
$F_{\eta\gamma\gamma}(0)$  we can obtain the $\eta-\eta'$ mixing parameters from Eqs.~(\ref{eq-1}-\ref{eq-4}). 
With three unknowns $(F_q,F_s,\phi)$, we must drop one of the equations. We discard the $\eta'$ asymptotics~\eqref{eq-4} since its determination, coming from the first ($N=1$) element, is the  
less reliable. Using the SL dataset alone, we obtained~\cite{Escribano:2013kba} 
\begin{equation}
\label{eq-6}
F_q/F_{\pi} =  1.06(1)\ , \ \ F_s/F_{\pi} = 1.56(24)\ , \ \ \phi = 40.3(1.8)^{\circ}\ ,
\end{equation}
where $F_{\pi}=92.21(14)$~MeV is the pion decay constant~\cite{Agashe:2014kda}. From the SL+TL dataset, we obtain~\cite{new}
\begin{equation}
\label{eq-7}
F_q/F_{\pi} =  1.07(1)\ , \ \ F_s/F_{\pi} = 1.39(14)\ , \ \ \phi = 39.3(1.2)^{\circ}\ ,
\end{equation}
which is a significant improvement compared to~\eqref{eq-6}. This translates to $F_8/F_{\pi} =  1.29(10),\  F_0/F_{\pi} = 1.19(6) , \ \theta_8 = -22.1(2.8)^{\circ} , \ \theta_0 = -8.1(3.2)^{\circ}$ in the 
octet-singlet basis~\cite{Escribano:2005qq,Feldmann:1998vh}. In Fig.~\ref{fig-4} we compare our determination~\eqref{eq-7} with different phenomenological results~\cite{Feldmann:1998vh,Escribano:2005qq} 
and find a very good agreement even though we use a much smaller amount of inputs than Refs.~\cite{Feldmann:1998vh,Escribano:2005qq,Escribano:2013kba}. 
Our determination is in tension with BaBar results 
at high TL $q^2$ values~\cite{Aubert:2006cy}. This urges for a second measurement of high-$Q^2$ data-points both for $\eta$ and $\eta'$ which may be accessed by the Belle collaboration. 

\section{Conclusions}

We have presented an easy and model-independent approach to describe and extract information of the $\eta$ and $\eta'$ transition form factors, such as the low energy parameters and the asymptotic 
behavior. Moreover, for the $\eta$ case, we have extended its application from the space-like region, where it was originally intended, to the time-like region, obtaining the most precise extraction of the 
low-energy parameters. Additionally we comment on the impact of the $\Gamma_{\eta\rightarrow\gamma\gamma}$ decay width on the slope parameter. Finally, we have used the information on the transition form 
factor to extract the $\eta-\eta'$ mixing parameters with an excellent compromise between precision and predictivity.

%

\begin{thebibliography}{15}

\bibitem{Adler:1969gk}
S.L. Adler, Phys.Rev. \textbf{177}, 2426 (1969); 
J.~Bell, R.~Jackiw, Nuovo Cim. \textbf{A60}, 47 (1969)

\bibitem{Lepage:1980fj}
G.P. Lepage, S.J. Brodsky, Phys.Rev. \textbf{D22}, 2157 (1980)

\bibitem{Escribano:2013kba}
R.~Escribano, P.~Masjuan, P.~Sanchez-Puertas, Phys.Rev. \textbf{D89}, 034014
  (2014), \texttt{1307.2061}

\bibitem{Feldmann:1998yc}
T.~Feldmann, P.~Kroll, Phys.Rev. \textbf{D58}, 057501 (1998),
  \texttt{hep-ph/9805294}

\bibitem{Feldmann:1998vh}
T.~Feldmann, P.~Kroll, B.~Stech, Phys.Rev. \textbf{D58}, 114006 (1998),
  \texttt{hep-ph/9802409}

\bibitem{Escribano:2005qq}
R.~Escribano, J.M. Frere, JHEP \textbf{0506}, 029 (2005),
  \texttt{hep-ph/0501072}

\bibitem{Masjuan:2008fv}
P.~Masjuan, S.~Peris, J.~Sanz-Cillero, Phys.Rev. \textbf{D78}, 074028 (2008),
  \texttt{0807.4893}

\bibitem{Masjuan:2012wy}
P.~Masjuan, Phys.Rev. \textbf{D86}, 094021 (2012), \texttt{1206.2549}

\bibitem{Baker}
G.A.Baker, P.~Graves-Morris, \emph{Pade Approximants} (Encyclopedia of
  Mathematics and its Applications, 1996)

\bibitem{Masjuan:2009wy}
P.~Masjuan, S.~Peris, Phys.Lett. \textbf{B686}, 307 (2010), \texttt{0903.0294}

\bibitem{Aguar-Bartolome:2013vpw}
P.~Aguar-Bartolome et~al. (A2 Collaboration), Phys.Rev. \textbf{C89}, 044608
  (2014), \texttt{1309.5648}

\bibitem{new}
R.~Escribano, P.~Masjuan, P.~Sanchez-Puertas, \emph{In preparation} ({})

\bibitem{Agashe:2014kda}
K.~Olive et~al. (Particle Data Group), Chin.Phys. \textbf{C38}, 090001 (2014)

\bibitem{Aubert:2006cy}
B.~Aubert et~al. (BaBar Collaboration), Phys.Rev. \textbf{D74}, 012002 (2006),
  \texttt{hep-ex/0605018}

\end{thebibliography}
%

\end{document}